# Real-time QoS Routing Scheme in SDN-based Robotic Cyber-Physical Systems

QoS Routing with SDN for Manufacturing Robotics

Rutvij H. Jhaveri
Delta-NTU Corporate Lab for CPS
Nanyang Technological University
Singapore
rhjhaveri@ntu.edu.sg

Rui Tan
School of Computer Science & Engg.
Nanyang Technological University
Singapore
tanrui@ntu.edu.sg

Sagar V. Ramani
Computer Engineering
Government Polytechnic
Porbandar, India
ramani_sagar@gtu.edu.in

***Abstract*— Industrial cyber-physical systems (CPS) have gained enormous attention of manufacturers in recent years due to their automation and cost reduction capabilities in the fourth industrial revolution (Industry 4.0). Such an industrial network of connected cyber and physical components may consist of highly expensive components such as robots. In order to provide efficient communication in such a network, it is imperative to improve the Quality-of-Service (QoS). Software Defined Networking (SDN) has become a key technology in realizing QoS concepts in a dynamic fashion by allowing a centralized controller to program each flow with a unified interface. However, state-of-the-art solutions do not effectively use the centralized visibility of SDN to fulfill QoS requirements of such industrial networks. In this paper, we propose an SDN-based routing mechanism which attempts to improve QoS in robotic cyber-physical systems which have hard real-time requirements. We exploit the SDN capabilities to dynamically select paths based on current link parameters in order to improve the QoS in such delay-constrained networks. We verify the efficiency of the proposed approach on a realistic industrial OpenFlow topology. Our experiments reveal that the proposed approach significantly outperforms an existing delay-based routing mechanism in terms of average throughput, end-to-end delay and jitter. The proposed solution would prove to be significant for the industrial applications in robotic cyber-physical systems.**

## I. INTRODUCTION

In order to improve the economy and fulfill the vision of smart manufacturing, most of the developed countries around the world have invested in the research towards Industry 4.0 [1]. Such smart manufacturing systems would be based on cyber-physical systems [2] where robotics would play a key role [3]. Such real-time systems demand stringent QoS requirements [4], [5]. Industrial communication systems, for the implementation of a robotic cyber-physical system (RCPS), demand strict delay guarantees of the flow along with high adaptability [6]. The fundamental function of the network layer in such a communication system is to determine a route from a source node to a destination node through a series of intermediate switches [7]. Traditional communication networks suffer from the inflexibility and inadaptability to the requirements of industrial environments [8]. Over the years, network operators have consistently attempted to improve network performance in order to fulfill application demands [9]. However, the complexity in achieving this goal kept increasing with the emergence of demanding applications until researchers were motivated to overcome these issues by Software-Defined Networking (SDN) paradigm which brought a novel concept of programmable networks for network managers.

SDN is a promising networking paradigm, which provides flexibility in network management and operation by providing global visibility and direct control of the forwarding elements. Thus, SDN has the capability to adapt to rapid changes in the network, which may prove to be vital for RCPS. In SDN, protocols such as OpenFlow provide flow level programmability in order to program the network according to QoS needs and network traffic conditions [10]. Industrial RCPS demands such QoS aware programming and reconfiguration, which assigns links based on their QoS needs and goals. However, there are several challenges that make QoS aware routing in RCPS a non-trivial problem: 1) real-time strict deadlines of the flows; 2) dynamic routing based on current network status and requirements. SDN controllers are capable of handling the complexity of route calculation and therefore, this critical function is offloaded to them [11]. Consequently, communication delay between the SDN controller and the network device becomes a crucial component, especially, for industrial applications [12].

In this work, we address the problem of discovering an efficient route based on current network status for delay-constrained RCPS by introducing a QoS aware Routing Scheme (QRS). Unlike most of the existing routing schemes which propose routing of new flows based on a single parameter (such as end-to-end delay), our scheme: 1) considers link parameters such as jitter, packet loss and link utilization to calculate link weight; 2) reroutes the new flows based on the current network state. In [13] and [14], jitter is mentioned as a crucial parameter in order to improve QoS in the industrial RCPS. The proposed scheme calculates link cost based on the aforementioned three factors and discovers an efficient route based on the link cost. This problem is challenging as the graph partitioning problem is an NP-hard problem [9], [15]. The experimental results show that the proposed scheme maximizes the service quality in comparison with an existing delay-based routing scheme.

The rest of the paper is organized as follows: Section 2 provides insight to the state-of-the art delay-constrained routing approaches for SDN. The details of our proposed delay-constrained QoS aware approach is discussed in Section 3. Section 4 describes our experimental setup and emulation results to show the effectiveness of our proposed approach. Finally, we conclude the paper in Section 5 along with future research directions.

## II. Related Work

A number of research approaches have been proposed for optimal routing traffic in SDN-based networks including industrial CPS. This section discusses the state-of-the-art in parameter-constrained routing in SDN.

Kumar *et al.* [4] proposed a mechanism which guarantees end-to-end delay requirements for high-criticality flows which estimates the requirements for flows into distinct queues based on *delay-monotonic policy*. It aims at satisfying the timing needs of hard real-time systems and at providing stable network performance even when different types of traffic are present in the system. However, the solution is complex in terms of establishment and maintenance of flow priorities. At the same time, it leads to the depletion of the available queues. Lee *et al.* [16] presented an energy-efficient heuristic mechanism based on segment routing which provides bandwidth guarantee. It attempts to reduce the unsatisfied request rate and improve the bandwidth satisfaction rate. The mechanism provides improved performance in terms of throughput and average rejection rate than other traditional routing mechanisms. A threshold-based model (TBM) proposed by Guck *et al.*[17] addresses QoS provisioning in real-time industrial networks by limiting the maximum delay of each queue. As a result, it does not need an a priori assignment of rate or buffer budgets. While TBM is flexible in automatically adapting to different types of traffic in the network, it increases the request processing time. Sudheera *et al.* [18] proposed a software defined vehicular network (SDVN) architecture which gives priority to essential delay needs in vehicular ad-hoc networks (VANETs). The proposed architecture does not exceed the latency of the existing VANET architectures, but still outperforms existing SDVN architectures as depicted by the theoretical and experimental comparisons.

Ozbek *et al.* [19] presented a low-complexity algorithm which satisfies throughput needs of all flows. The scheme provides energy efficient routing by limiting the energy consumption. However, the scheme assumes low network load and less number of flows. Looking at the industrial networks that require hard real-time guarantees, Tomovic *et al.* [11] presented a solution for delay-constrained SDN networks which classifies the flows according to their delay sensitivity. Additionally, it attempts to reduce computational complexity and to maximize resource utilization. Routes are calculated only when network is initialized and re-calculated when topology changes. However, the solution adopts Yen's algorithm [20] to calculate important paths which gives high degree of overlapping.

Most of the aforementioned works propose flexible network management schemes while serving the sole purpose of selection of an optimal route based on a single or multiple constraints. They do not address the vital issue of improving overall QoS of a delay-constrained network such as RCPS. In addition, few schemes have considered the all-important issue of dynamic routing during link failure or link status change.

## III. System Model

We consider an SDN-based industrial network, which uses OpenFlow protocol for dynamic programming of the network. The flows are forwarded by switch according to its forwarding table with the objective of decoupling network intelligence and forwarding element. The controller can obtain the network state by querying the switches. The goal of the proposed algorithm is to discover an efficient route based on the cost of each link.

The proposed QoS aware Routing Scheme (QRS) computes all possible paths of the network between the two hosts and selects the optimal path for routing based on different QoS parameters such as jitter, packet loss and link utilization. As SDN controller has a global view of the topology and it collects the QoS statistics from the OpenFlow switches. As it is complex and expensive to measure link latency using the existing approaches, we exploit the LLDP protocol (a vendor neutral layer 2 protocol) to serve this purpose. LLDP can be used to measure link latency with minimal overhead with the accuracy of milliseconds (as proposed in [21]). In order to reduce the traffic overhead in network and to improve the accuracy of measurement in real-time, inherent packets of OpenFlow SDN are used (such as LLDP, Packet-Out, Packet-In and Echo messages). LLDP packet is modified to include the timestamp information in one of its type-length-value (TLV) structures. The working of Echo messages and the procedures of link discovery to measure link delay with LLDP for a single link SDN network are presented in Figure 1. The reverse link is discovered in a similar way. For collecting the flow related statistics *FLOWSTATS_REQUEST* and *FLOWSTATS_REPLY* messages are used [22].

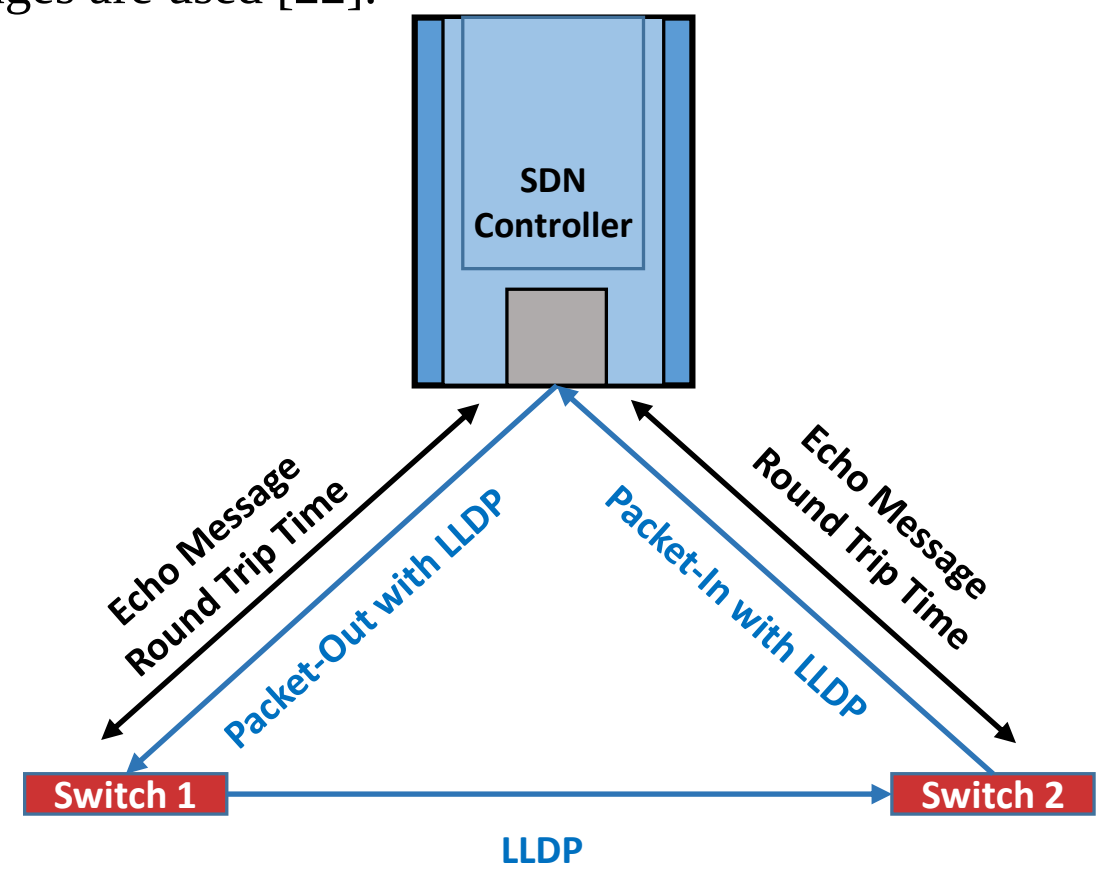

Figure 1. Link discovery procedures with LLDP

Figure 2 represents the architecture of our proposed scheme, QRS. The *QoS Calculator Module* calculates the QoS cost of each link based on the statistics supplied by *Delay Detector Module* and *Flow Collector Module*. The routing algorithm calculates the shortest-path based on Dijkstra's algorithm by taking each link's QoS cost as the link weight. Finally, the flows are forwarded on the calculated shortest-path. Meanwhile, the controller diverts the flows to an alternative path based on the computed cost metric during the cost update interval. Thus, the traffic is diverted to an alternative path during unwanted network conditions.

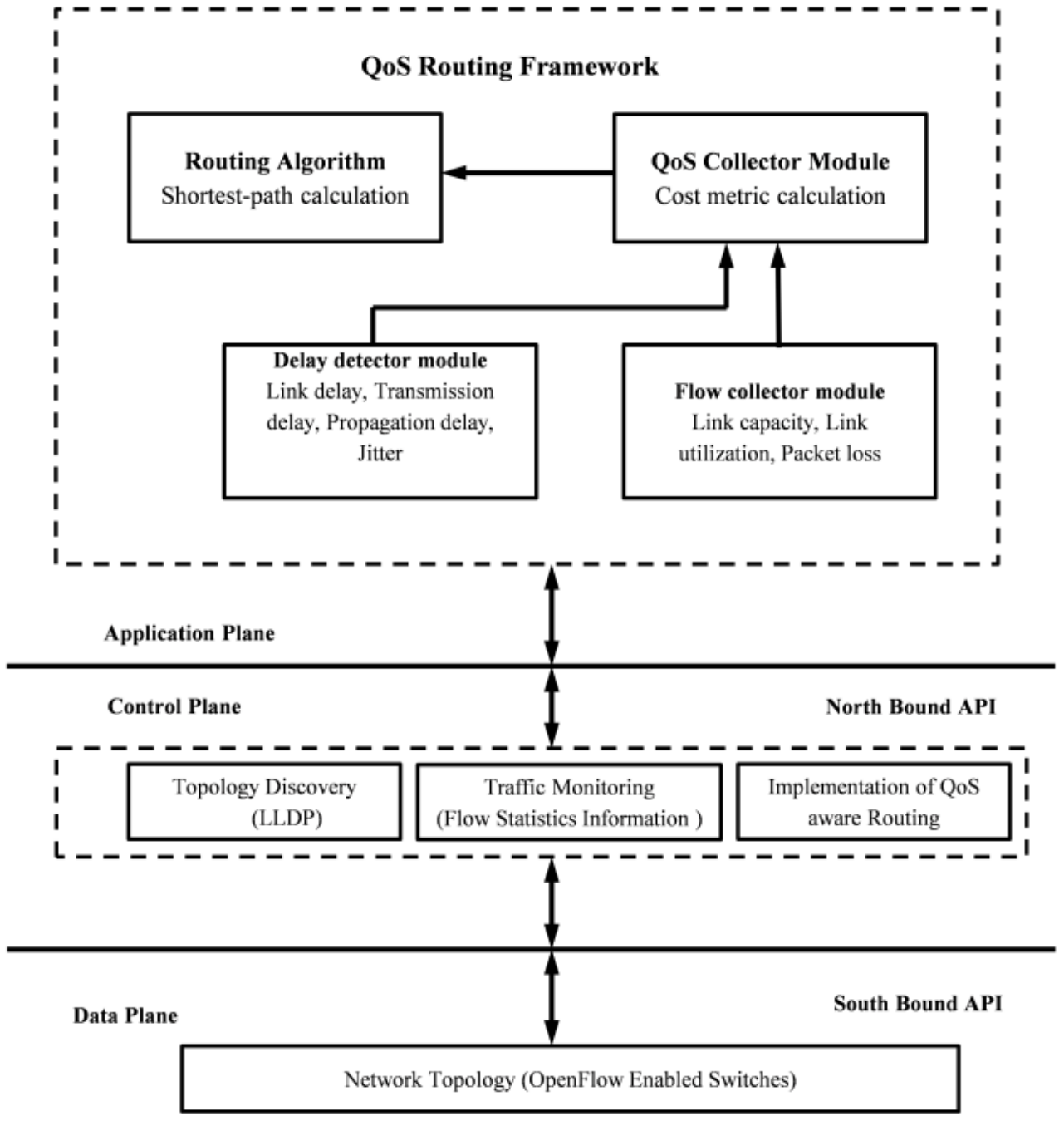

Figure 2: SDN Architecture for QoS-aware Routing Scheme

QRS discovers an optimal path, in terms of QoS, between a source and a destination if the path satisfies the delay requirements of the flow. Consider: 1) the network topology as a graph *G*; 2) the number of nodes in the network as *V*; 3) $P = \{p_1, p_2, p_3, ..., p_N\}$ as *N* possible number of edge-disjoint paths from the given source (*src*) to the destination (*dst*); 4) $p_j$ as a set of *m* edges between *m+1* nodes of the given graph *G* with the corresponding delay $D_j = \{d_{j1}, d_{j2}, d_{j3}, .., d_{jm}\}$ and; 5) $TD_j$ as the end-to-end delay of path the $p_j$. In order to compute the delay of each intermediate link of $p_j$, $d_{jk}$ (where $1 \leq k \leq$ m), the controller sends an LLDP message to the directly connected switches. The link delay, $d_{jk}$, can be obtained by equation (1) [21]:

$$d_{jk} = (t_i + t_{i+1} - latency_i - latency_{i+1})/2 \quad (1)$$

where $t_i$ is the traversal time for LLDP to traverse from the controller to node *i*, node *i* to node *i+1* and, node *i+1* back to the controller ($t_{i+1}$ is calculated in the similar way); $latency_i$ and $latency_{i+1}$ are the times consumed by the echo messages from the controller to nodes *i* and *i+1* respectively.

Thus, delay $D_j$ can be computed by equation (2):

$$D_j = \sum_{i=1}^{m} d_{ji} \quad (2)$$

The transmission delay ($d_{trans}$) from a node *i* can be conceptualized as per below equation (3):

$$d_{trans_i} = \frac{packet\ length}{link\ bandwidth} \quad (3)$$

Consequently, the corresponding end-to-end delay can be measured for the given path $p_j$ using equation (4):

$$TD_j = D_j + \sum_{i=1}^{m+1} d_{trans_i} \quad (4)$$

QRS computes QoS cost of a path $p_j$ ($C_{pj}$) if the path satisfies the end-to-end delay guarantees. The cost of a path $p_j$ is calculated with equation (5):

$$C_{p_j} = \sum_{i=1}^{m+1} \left(w1 * \frac{jitter_{i,i+1}}{average\ jitter_{i,i+1}} + w2 * \frac{packets\ droped_{i,i+1}}{packets\ sent_{i,i+1}} + w3 * \frac{link\ utilization_{i,i+1}}{link\ capacity_{i,i+1}}\right) \quad (5)$$

where $jitter_{i,i+1}$ and average $jitter_{i,i+1}$ are the current jitter and average jitter calculated for the link between *i* and *i+1* nodes, respectively; ($packets\ dropped_{i,\ i+1}$ / $packets\ sent_{i,\ i+1}$) is the packet drop ratio of the link; link $utilization_{i,i+1}$ and link $capacity_{i,\ i+1}$ are the utilization and total bandwidth of the link between the aforementioned nodes; *w1*, *w2* and *w3* are the weights given to each of these three factors. Dijkstra's shortest path algorithm is employed to discover the shortest path from source to destination as per link weights calculated with equation (5) when a new flow arrives. The new flow is then routed through this cost effective path.

## IV. EMULATION RESULTS

In what follows, we present the set-up of our simulations. In order to evaluate the performance of our scheme, different performance metrics are used to compare its performance with an existing scheme.

### A. *Emulation Setup and Tools*

Table I summarizes the tools used for implementation of our scheme and experimental evaluation.

TABLE I. TOOLS USED FOR EXPERIMENTS

| Software and Version | Function |
|---|---|
| Ubuntu 16.04 | Host operating system |
| Mininet 2.3.0d4 | Network emulator |
| OpenFlow 1.3 | SDN protocol for southbound interface |
| Ryu 4.26 | SDN Controller |
| Python 2.7.12 | Programming language |
| Iperf 3.1.3 | Generating traffic |

In the experiments, we consider: 1) a realistic industrial system [23] as shown in Figure 3; 2) all flows to be TCP

flow; 3) all of the wired links are of 100 meters and have 1 Gbps link capacity; 4) 4 source-destination pairs are selected randomly; 5) each flow is sent at 200 Mbps rate; 6) *w1=w2=w3=1/3*. During the route migration phase, along with its routing information, the controller updates the forwarding tables of switches with the use of OpenFlow messages to each switch. The frequency of route migration depends on the network status while a new flow arrives [24].

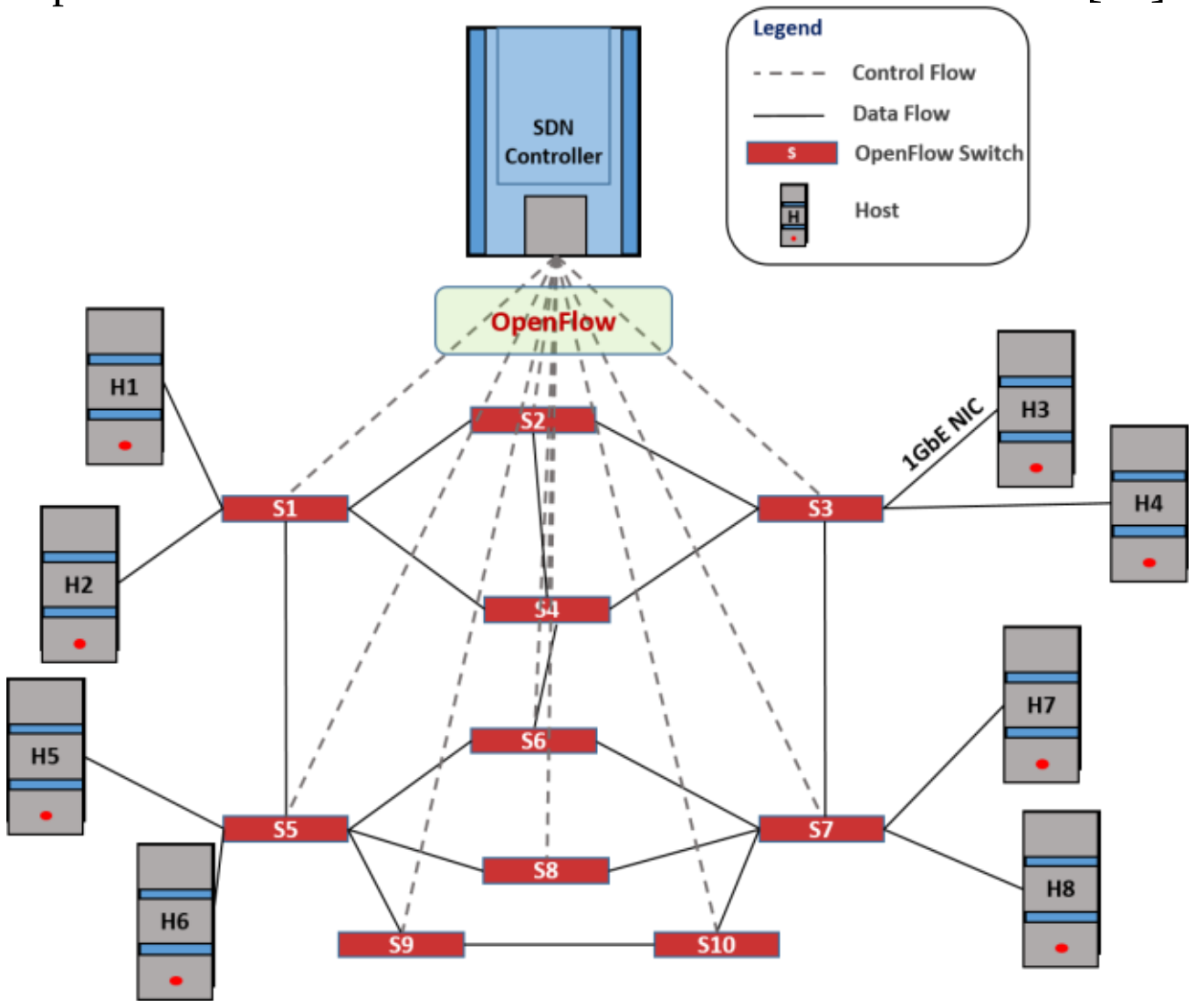

Figure 3: Mininet Testbed

### B. *Results and Performance Analysis*

In this subsection, we graphically compare the performance of the proposed scheme (QRS) with an existing delay-based shortest-path routing scheme (LLMP) presented in [21] in the congested environment.

*1) Test 1: Varying Number of Flows.* In this test, we evaluate the performance of the proposed scheme, QRS, by varying the number of flows from 1 to 5 with emulation time of 300 sec.

As shown in Figure 4, as the number of flows increases the average jitter increases due to the network congestion. Meanwhile, the average jitter of QRS stays below to that of the LLMP routing scheme in all the cases. The average jitter of QRS is 0.162 msec which is significantly less than 0.220 msec of LLMP. This is because QRS selects the path with lesser jitter.

As shown in Figure 5, as the number of flows increases the average end-to-end delay increases as the packets suffer from delays due to increased network congestion. Meanwhile, the average end-to-end delay of QRS is 2.447 msec which is significantly lower than 2.924 msec of LLMP. This is because QRS selects path with higher available bandwidth as well as lower packet loss.

As shown in Figure 6, it is obvious that as the number of flows increases, the average network throughput increases for both the schemes. The average throughput provided by QRS is 583.63 Mbps while that of LLMP is 562.22 Mbps. This is because QRS takes into account delay, bandwidth and packet loss while selecting the path.

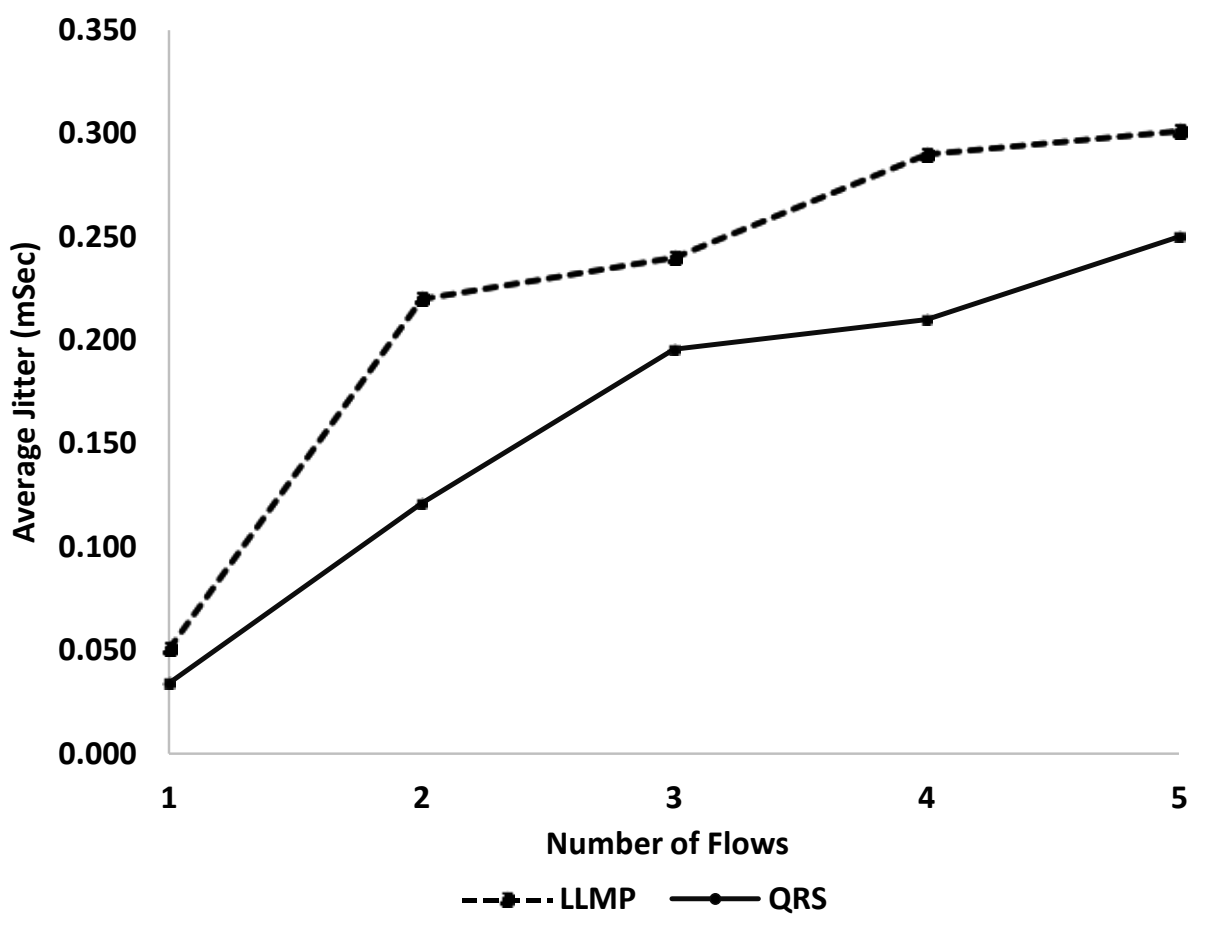

Figure 4: Average Jitter Vs Number of Flows

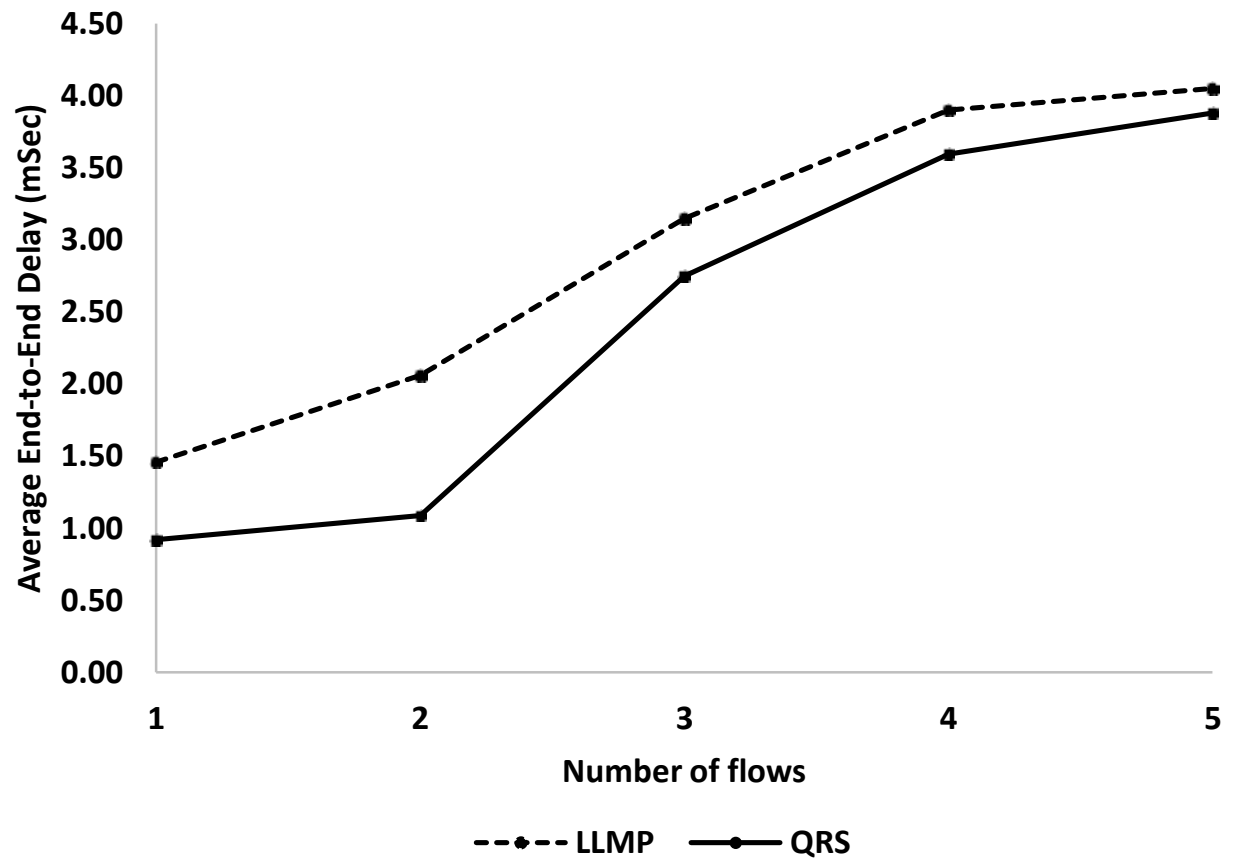

Figure 5: Average Delay Vs Number of Flows

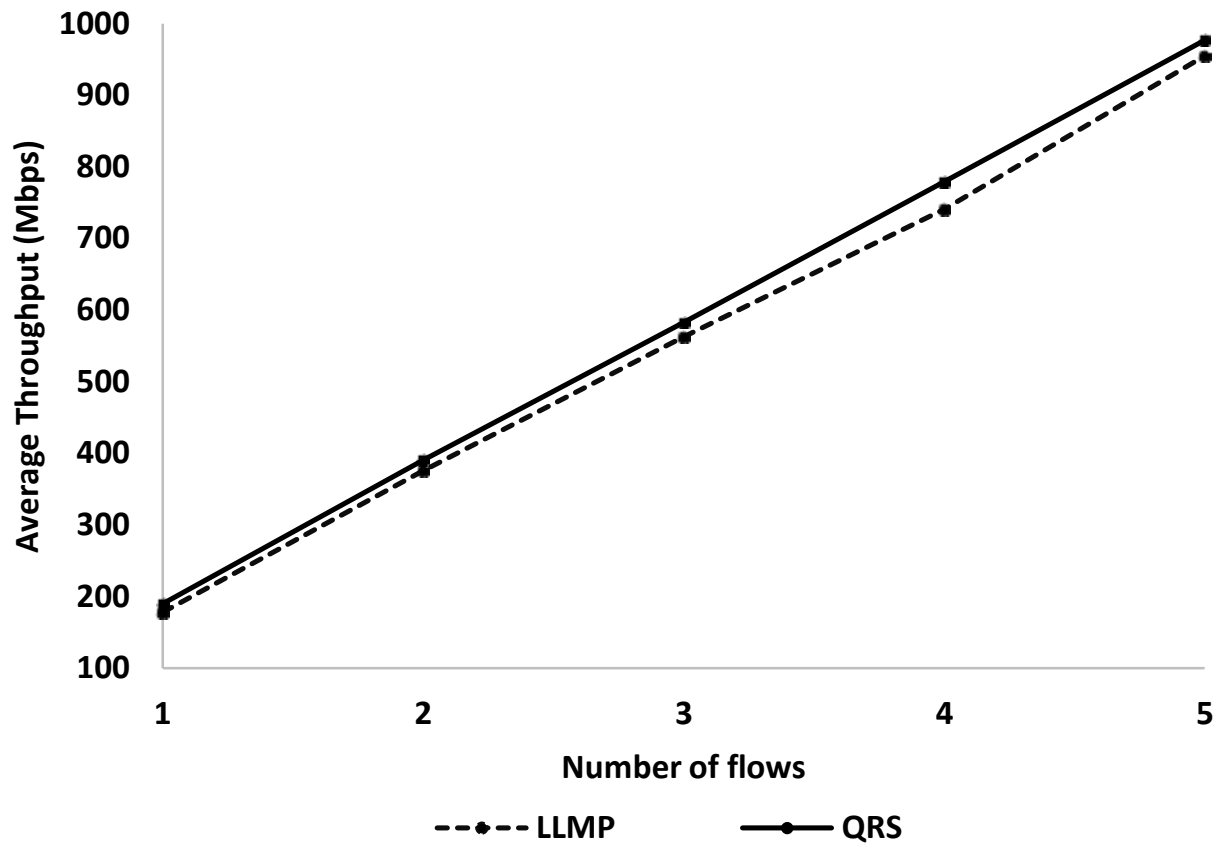

Figure 6: Average Throughput Vs Number of Flows

Thus, in this test, QRS improves average jitter by 26.36%, average end-to-end delay by 16.31% and average throughput by 3.81%.

*2) Test 2: Varying Time.* In this test, we evaluate the performance of QRS by varying the emulation time from 0 to 300 seconds and taking the number of flows as 2.

As shown in Figure 7, due to the aforementioned reason, the average jitter of QRS is 0.121 msec while that of LLMP 0.243 is msec which is significantly higher than that of QRS.

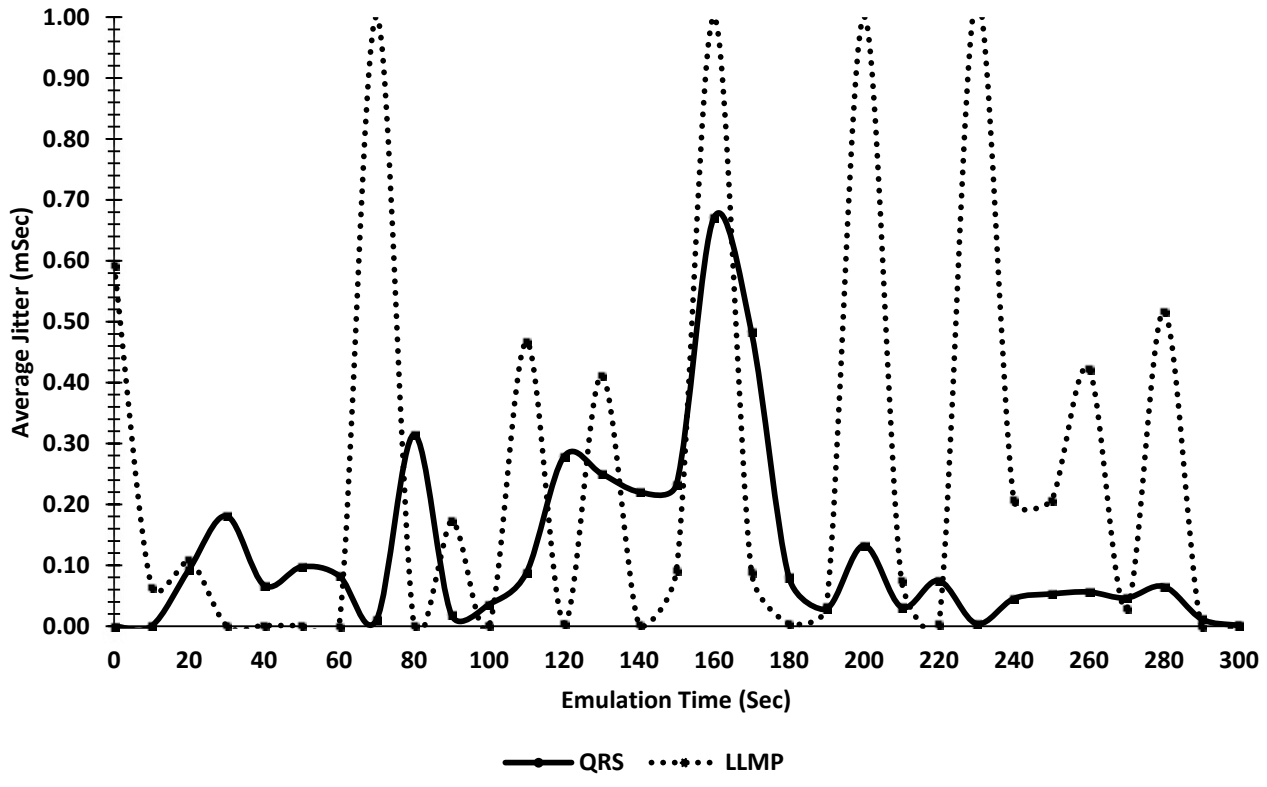

Figure 7: Average Jitter Vs Emulation Time

As shown in Figure 8, due to the aforementioned reasons, the average end-to-end delay of QRS is 1.083 msec while that of LLMP is 2.054 msec which is considerably higher than that of QRS.

As the results depict, in this test, QRS improves average jitter by 50.21% and average end-to-end delay by 47.27%. Thus, we can state that our proposed approach performs significantly better in the delay-constrained environment than LLMP.

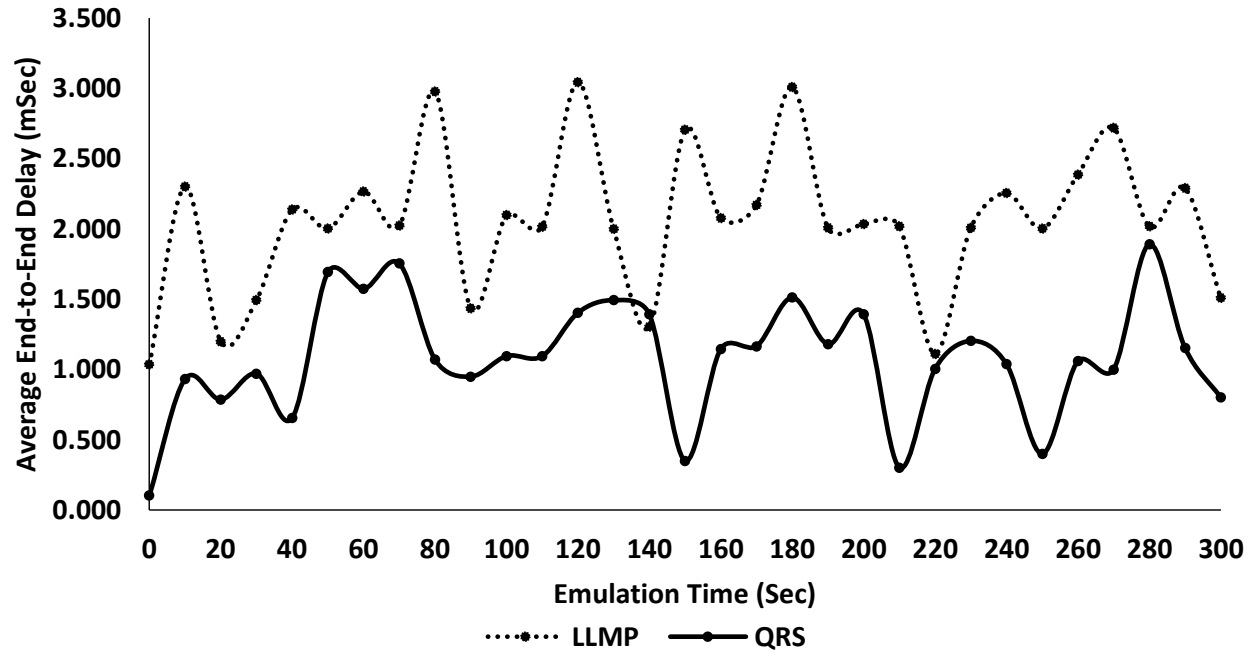

Figure 8: Average Delay Vs Emulation Time

## V. Conclusion

In this paper, we address the problem of QoS routing in delay-constrained robotic cyber-physical system. We primarily focus on improving quality-of-services for delay-constrained flows, and moving a step forward, we attempt to migrate routes dynamically based on the current network status. The proposed scheme dynamically discovers a QoS efficient route with minimal communication overhead while continuously monitoring the network links with multi-dimensional cost metric. In order to provide resiliency in the network, the scheme reacts to the abnormal network state by adopting a strategy to migrate the flows to more stable alternative routes. The experimental results depict that QRS outperforms the existing scheme in terms of average jitter, average end-to-end delay and average throughput. We state that QRS can provide significant improvement in the quality-of-services in robotic CPS. In future, as a part of resiliency research in industrial CPS, we plan to devise a resilient reactive mechanism which can provide delay guarantees during link failures and high congestions.

## Acknowledgment

This research work was conducted within the Delta-NTU Corporate Laboratory for Cyber-Physical Systems with funding support from Delta Electronics Inc. and the National Research Foundation (NRF), Singapore under the Corp Lab @ University Scheme. We would like to acknowledge the contributions from Prof. Arvind Easwaran, NTU, Singapore and Sidharta Andalam, Delta Electronics, Singapore for their valuable inputs.